\pdfoutput=1

\documentclass[aps,prd,amsmath,amssymb,showpacs,twocolumn,superscriptaddress,sort&compress]{revtex4-1}

\usepackage{amsmath}
\usepackage{amssymb}
\usepackage{graphicx}
\usepackage{dsfont}
\usepackage{units}
\usepackage{bm}
\usepackage{slashed}
\usepackage[colorlinks=true,linkcolor=blue,citecolor=blue,urlcolor=blue]{hyperref}

         \newcommand{\be}{\begin{equation}}
         \newcommand{\ee}{\end{equation}}

         \newcommand{\abs}[1]{\ensuremath{\left\lvert#1\right\rvert}}

         \newcommand{\one}{\ensuremath{\mathbf{1}}}

         \newcommand{\conjg}[1]{\ensuremath{\hspace{1pt}\overline{\hspace{-1pt}#1\hspace{-1pt}}}\hspace{1pt}}
         \newcommand{\vect}[1]{\bm{#1}}

         \newcommand{\ket}[1]{\ensuremath{\vert #1 \rangle}}

         \renewcommand{\sp}{\ensuremath{\mathrm{u}}}

\begin{document}

         \title{Hadronic decays of mesons and baryons in the Dyson-Schwinger approach}

         \author{V. Mader}
          \email{valentin.mader@uni-graz.at}
          \affiliation{Institut f\"ur Physik, Karl-Franzens-Universit\"at Graz, A-8010 Graz, Austria}

         \author{G. Eichmann}
         \affiliation{Institut f\"ur Theoretische Physik I, Justus-Liebig Universit\"at Giessen, D-35392 Giessen, Germany}

         \author{M. Blank}
          \affiliation{Institut f\"ur Physik, Karl-Franzens-Universit\"at Graz, A-8010 Graz, Austria}

         \author{A. Krassnigg}
          \affiliation{Institut f\"ur Physik, Karl-Franzens-Universit\"at Graz, A-8010 Graz, Austria}

         \date{\today}

         \begin{abstract}
         We study hadronic decays of mesons and baryons in the context of the
         Dyson-Schwinger equations of QCD. Starting from a well-established
         effective interaction in rainbow-ladder truncation, we consistently
         calculate all ingredients of the appropriate decay diagrams.
         The resulting strong couplings are presented as functions of the quark mass
         from the chiral limit up to the respective decay thresholds.
         In particular, we investigate the $\rho\to\pi\pi$ and for the
         first time the $\Delta\to N\pi$ transitions. Both meson and baryon results
         compare well to available lattice QCD results as well as experimental data
         and present the first step towards a comprehensive covariant study of
         hadron resonances in the Dyson-Schwinger approach.
         \end{abstract}


         \pacs{%
         %
         %
         %
         %
         11.10.St 
         12.38.Lg, 
         13.25.-k  
         13.30.Eg  
         %
         %
         }
         \maketitle


\section{\label{sec:intro} Introduction}

         In hadron physics the strong interaction dominates the decay width of a
         resonance if appropriate hadronic channels are open. Thus, strong decays
         are an issue of paramount interest and any study of hadrons as mere bound
         states is necessarily incomplete if the states under investigation lie
         above the relevant thresholds.

         Consequently, strong decays of hadrons in terms of composites of
         quarks and gluons have been studied from the beginning, e.g., in
         quark models, and various decay mechanisms, formulations, and levels
         of sophistication have been employed over the years. We exemplarily mention harmonic-oscillator models
         \cite{Faiman:1969at,Feynman:1971wr,Kim:1974vj}, the elementary-emission model
         \cite{Koniuk:1979vy,Godfrey:1985xj,Plessas:1999nb,Melde:2006yw},
         flux-tube breaking~\cite{Kokoski:1985is,Close:1994hc,Stassart:1995qf}, the
         quark-pair creation or ${}^3P_0$ mechanism~\cite{LeYaouanc:1974mr,Capstick:1998uh,Theussl:2000sj}
         and a simple coupled-channel formulation of meson resonance properties in relativistic
         Hamiltonian dynamics \cite{Krassnigg:2003bd,Krassnigg:2003gh,Biernat:2010tp,Kleinhappel:2010hh}.
         These approaches range from the aforementioned non- and semirelativistic as well as fully
         Poincar\'{e} invariant quark-model calculations with various interactions
         to reductions of a Bethe-Salpeter treatment of hadrons, e.g.,~\cite{Singh:1988vk,Metsch:2003ix,Ricken:2003ua,Guo:2007qu}.
         Ideally one would have a consistent and comprehensive coupled-channel calculation of meson and baryon
         resonances, see e.g.~\cite{Oset:2010ab}.
         However, to consistently implement an analogous approach within QCD is challenging.
         Recent lattice studies provide promising progress
         \cite{Gockeler:2008kc,Feng:2009ck,Feng:2010es,Aoki:2010hn,Frison:2010ws,Lang:2011mn,Alexandrou:2010uk};
         nevertheless it is encouraging that reasonable results can already be obtained
         without working with resonances from the very beginning.

         In the Dyson-Schwinger-equation (DSE) approach to QCD two-body systems are described
         by the Bethe-Salpeter equation (BSE). Hadronic transition processes were considered
	 in \cite{Pichowsky:1996tn,Pichowsky:1999mu}, and an exploratory study of the strong
         decay of light vector mesons to two pseudoscalars was performed in this framework
	 some years ago \cite{Jarecke:2002xd}.
         Here we revisit this study and investigate in detail both model-parameter
         and quark-mass dependence of the results, and we compare to experimental data
         and analogous calculations in lattice-regularized QCD.

         As the next important step towards a comprehensive investigation
         of the entire hadron spectrum, we generalize this calculation to a strong
         baryon decay and calculate the hadronic coupling of the $\Delta-$baryon to $N\pi$.
         The $\Delta(1232)$ resonance plays an important role in pion-nucleon scattering and
         pion photoproduction experiments; see~\cite{Cattapan:2002rx,Pascalutsa:2006up} for recent reviews.
         Its experimental width mainly owes to the decay into a nucleon and a pion,
         whereas the electromagnetic $\Delta\rightarrow N\gamma$ decay channel is considerably suppressed.
         The corresponding strong coupling constant $g_{\text{\tiny{$\Delta N$}}\pi}$ has been studied in various approaches, e.g.,
         in the quark model~\cite{Niskanen:1981pq,Melde:2008dg}, in meson-exchange models~\cite{Sato:1996gk,Polinder:2005sn},
         via light-cone sum rules~\cite{Wang:2007em,Aliev:2010su}, and recently also in lattice QCD~\cite{Alexandrou:2008bn}.
         The $\Delta-$baryon is an essential component in a realistic description of meson-baryon interactions
         in nuclear physics, for instance via chiral effective field theories,
         and the computation of the $\Delta N\pi$ vertex from its underlying quark-gluon dynamics
         is an important task in hadron physics.

         The present study is motivated by recent successes in the implementation
         of the Dyson-Schwinger approach to various aspects of hadron phenomenology.
         In particular we make use of a well-established effective model setup
         in rainbow-ladder (RL) truncation, together with the consistent extension to
         a quark-diquark picture for baryons.
         Successful applications of this setup pertain to various observables:
         meson spectra and leptonic decay constants of pseudoscalar and vector mesons were studied
         over the full range of quark masses from chiral limit to bottomonium \cite{Maris:2006ea,Krassnigg:2009zh},
         and recently the feasibility of such meson studies for any spin has been demonstrated \cite{Krassnigg:2010mh}.
         Electromagnetic properties of pseudoscalar and vector mesons
         involve a consistent construction of the electromagnetic interaction process
         via triangle diagrams analogous to the one used here~\cite{Maris:1999bh,Maris:2000sk,Holl:2005vu,Bhagwat:2006pu}.
         They have proven to be an excellent example for the importance of
         correctly implementing the symmetry properties of the underlying
         theory in numerical calculations.

         The extension of the approach to baryons can be simplified via the introduction of
         diquarks \cite{Anselmino:1992vg,Maris:2002yu,Maris:2004bp}. The corresponding investigations of baryons
         in a covariant quark-diquark setup have already undergone considerable development
         \cite{Oettel:1998bk,Oettel:2000jj,Eichmann:2007nn,Eichmann:2008ef,Cloet:2008re,Nicmorus:2008vb,Nicmorus:2010sd}
         and very recently culminated in the first genuine three-quark treatment of
         nucleon and $\Delta$ masses and nucleon electromagnetic form factors
         in the covariant Faddeev framework~\cite{Eichmann:2009zx,Eichmann:2009qa,SanchisAlepuz:2010in,Eichmann:2011vu}.

         This article is organized as follows: in Section~\ref{sec:bb} we collect
         the necessary building blocks of the approach;
         the $\rho\rightarrow\pi\pi$ and $\Delta\rightarrow N\pi$ transition diagrams are worked out in Section~\ref{sec:decays};
         Section~\ref{sec:results} contains our results
         for both meson and baryon sectors; and we conclude in Section~\ref{sec:conclusions}.
         Decay-width formulas and the color-flavor traces of the decay diagrams are collected in the appendices.
         All calculations are performed in Euclidean momentum space and in
         the isosymmetric limit in Landau gauge QCD.

\section{\label{sec:bb}Building Blocks}

         Our investigation of hadronic decays involves numerical
         solutions of several integral equations whose properties and solutions
         have been studied elsewhere. In particular, we are concerned with
         the quark DSE, the meson and diquark BSEs and, in the context of baryons, the quark-diquark BSE.
         In the following we briefly
         review these equations and the properties of their solutions, and
         for each case we refer the reader to more detailed discussions in the literature.

        \subsection{\label{sec:interaction}Truncation and effective interaction}

         Numerical model studies of hadrons such as the one presented here necessitate a
         truncation of the infinite tower of DSEs. In the following we will restrict ourselves
         to the RL truncation which substitutes the fully dressed quark-gluon vertex with a bare vertex.
         Its counterpart in a hadronic bound-state equation is a gluon ladder kernel, i.e.
         a dressed iterated gluon exchange between two quarks.
         The combined strength of the gluon propagator and quark-gluon vertex
         is then modeled by an ansatz.
         Phenomenologically important directions of improvement beyond RL involve the implementation of pseudoscalar meson-cloud effects
         but also other structures in the quark-gluon vertex and the $qq$ and $q\bar{q}$ kernels,
         see~\cite{Alkofer:2008et,Williams:2009ce,Chang:2010jq} and references therein.

         The RL truncation offers many
         advantages. It is simple to implement, but at the same time allows for sophisticated
         model approaches to QCD within the DSE-BSE context since it satisfies the axial-vector
         and vector Ward-Takahashi identities
         (see e.g.~\cite{Maskawa:1975hx,Aoki:1990yp,Kugo:1992zg,Bando:1993qy,Munczek:1994zz,Maris:1997hd,Maris:1999bh,Maris:2000sk}).
         The axial-vector Ward-Takahashi identity is essential for the correct realization
         of chiral symmetry and its dynamical breaking in any model calculation.
         In particular, it imposes constraints upon the construction of the integral-equation kernels.
         As the most prominent result, Goldstone's theorem
         is satisfied~\cite{Munczek:1994zz} and one obtains a generalized Gell-Mann--Oakes--Renner relation valid
         for all pseudoscalar mesons and all current-quark masses~\cite{Maris:1997tm,Holl:2004fr}.
         The vector Ward-Takahashi identity on the other hand is the guiding principle for the
         construction of consistent electromagnetic currents.

         In RL truncation the equations as presented in the following subsections contain an essential model ingredient,
         namely an effective interaction $\mathcal{G}$ which we choose from Ref.~\cite{Maris:1999nt} as
         \begin{equation}\label{eq:interaction}
         \frac{{\cal G}(k^2)}{4\pi Z_2^2} = \frac{D\pi \,k^4}{\omega^6} \,\mathrm{e}^{-k^2/\omega^2}
         +\frac{2\pi\gamma_m \,(1 - e^{-k^2/\Lambda_t^2}) }{\ln \,[\tau+(1+k^2/\Lambda_\mathrm{QCD}^2)^2]}\,,
         \end{equation}
         where $k$ is the gluon momentum and $Z_2$ the quark renormalization constant.
         This particular form has been employed in many of the works listed as successes of the approach in the
         introduction. It provides the correct amount of dynamical chiral symmetry breaking as well as quark confinement
         via the absence of a Lehmann representation for the dressed quark propagator. Furthermore, it has the
         correct perturbative limit, i.\,e.~it preserves the one-loop renormalization group behavior of QCD for
         solutions of the quark DSE. Following~\cite{Maris:1999nt}, we have $\Lambda_t=1$~GeV,
         $\tau={\rm e}^2-1$, $N_f=4$, $\Lambda_\mathrm{QCD}^{N_f=4}= 0.234\,{\rm GeV}$, and $\gamma_m=12/(33-2N_f)$.
         The main motivation of this function, which mimics the behavior of the product of quark-gluon vertex and gluon propagator, is
         of phenomenological origin. While currently debated on principle grounds (e.g.~\cite{Fischer:2008uz,Binosi:2009qm})
         the impact of its particular form in the far IR on meson masses is expected to be small
         (see also \cite{Blank:2010pa} for an exploratory study in this direction).

         The phenomenologically important regime, in particular with respect to light meson
         (e.g., pion) properties, is the intermediate-momentum region, modeled by the Gaussian term in Eq.~\eqref{eq:interaction}.
         $D$ and $\omega$, in principle free parameters of the model interaction, can be used to investigate
         certain aspects of both the interaction and the bound states in the BSE. In particular one
         can interpret $D$ as an overall strength and $\omega$ as an inverse effective range of the interaction.
         In the range $\omega\in [0.3,0.5]$ GeV, the prescription $D\times\omega=:\Lambda_\text{IR}^3=const.$ follows from
         fitting of the model parameters to ground-state properties~\cite{Maris:1999nt} and defines a one-parameter model,
         characterized by a fixed infrared scale $\Lambda_\text{IR}=0.72$ GeV.
         Apparently~\cite{Krassnigg:2009zh} an insensitivity of an observable to $\omega$ in this
         prescription is characteristic for a ground state, while orbitally or radially excited-state properties
         show considerable dependencies on $\omega$. In the present work, we also consider such a possibility and
         plot bands to indicate the model-parameter dependence of our results.

         \begin{figure*}[t]
                    \begin{center}

                    \includegraphics[scale=0.13]{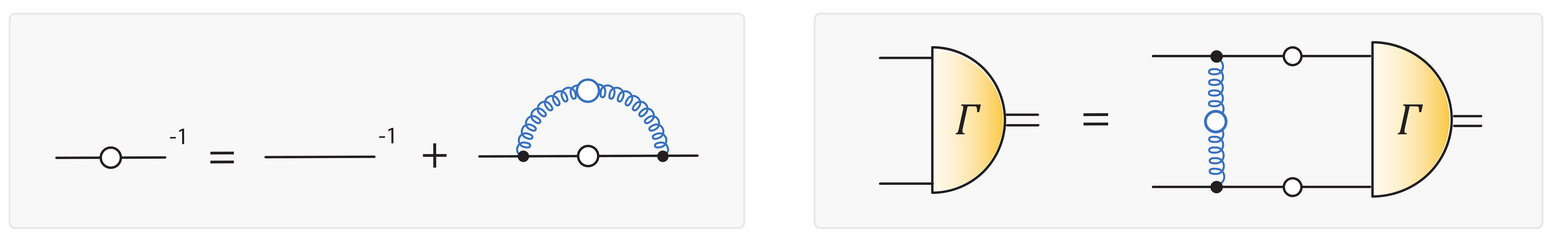}
                    \caption{(Color online) \textit{Left panel:} quark DSE~\eqref{eq:dse} in rainbow truncation.
                              \textit{Right panel:} schematics of the meson BSE~\eqref{eq:bse} and diquark BSE~\eqref{eq:diBSEtr} in ladder truncation. } \label{fig:DSE-BSE}

                    \end{center}
        \end{figure*}

        \subsection{\label{sec:quarks}Quarks}

         The Dyson-Schwinger equation for the quark propagator in rainbow
         truncation is illustrated in the left panel of Fig.~\ref{fig:DSE-BSE} and reads
         \begin{equation}\label{eq:dse}
         S(p)^{-1}  =  Z_2\,(i\slashed{p} + m_0)+  \frac{4}{3}\int\limits_q  \mathcal{G}(k^2) \, \frac{T_k^{\mu\nu}}{k^2}\,
         \gamma^\mu \,S(q)\, \gamma^\nu \,,
         \end{equation}
         where $S(p)$ is the renormalized dressed quark propagator, $p$ and $k=p-q$ are the quark and gluon momenta,
         and $\int_q = \int d^4q/(2\pi)^4$ represents a four-momentum integration.
         $T^{\mu\nu}_k/k^2$ with
         $T^{\mu\nu}_k=\delta^{\mu\nu}-k^\mu k^\nu/k^2$ is the free gluon
         propagator, $\gamma^\nu$ is the bare quark-gluon vertex, and $\mathcal{G}(k^2)$
         is the effective interaction defined above in Eq.~(\ref{eq:interaction}).
         Dirac and flavor indices have been omitted for simplicity and the factor
         $\nicefrac{4}{3}$ comes from the color trace.
         The bare current-quark mass $m_0$ is the input of the equation.
         The solution of Eq.~(\ref{eq:dse}) requires a renormalization procedure, the details of which
         can be found together with the general structure of the quark DSE in e.g., \cite{Maris:1997tm,Maris:1999nt}.

         $S(p)$ is an important ingredient in all of the following. We note
         that its solution as an input for the various BSEs described below must be
         known in a parabola-shaped region of the complex $p^2$ plane, the size of which is proportional to the mass
         of the respective bound state that it helps to constitute. As a result, a sophisticated
         numerical approach is needed and we refer the reader to \cite{Krassnigg:2008gd} for a
         description of our particular solution method.

        \subsection{\label{sec:mesons}Mesons}

         In our framework mesons with total $q\bar{q}$ momentum $P$ and relative $q\bar{q}$
         momentum $p$ are studied via the meson BSE.
         Its structure in RL truncation is shown in the right panel of Fig.~\ref{fig:DSE-BSE} and reads
         \begin{equation}\label{eq:bse}
         \Gamma_\text{M}(p,P)=-\frac{4}{3}\int\limits_q \mathcal{G}(k^2)\, \frac{T_k^{\mu\nu}}{k^2} \gamma^\mu   \chi_\text{M}(q,P) \,\gamma^\nu \,,
         \end{equation}
         where $\Gamma_\text{M}(k;P)$ is the Bethe-Salpeter amplitude and
         $\chi_\text{M}(q;P)=S(q_+) \,\Gamma_\text{M}(q;P) \,S(q_-)$ is referred to as the
         Bethe-Salpeter wave function.
         The quark and antiquark propagators
         depend on the (anti)quark momenta $q_+ = q+\eta P$ and $q_- = q- (1-\eta) P$, where $\eta \in [0,1]$ is
         a momentum partitioning parameter usually set to $1/2$ for systems of equal-mass constituents (which we do as well).
         The quark-antiquark interaction kernel
         is given by a ladder dressed-gluon exchange, whose dependence on the gluon momentum is characterized
         by the same effective interaction, Eq.~(\ref{eq:interaction}), as in the quark DSE Eq.~(\ref{eq:dse}).
         The combined set of truncated equations (\ref{eq:dse}) and (\ref{eq:bse}) thus already by construction
         satisfies the axial-vector Ward-Takahashi identity.

         The general dependence of the meson amplitude on the various four-momenta can be written in terms of
         $N$ covariant structures $T_i$ reflecting the quark and antiquark spins,
         together with scalar components $F_i$, $i=1,\ldots, N$ as
         \begin{eqnarray}\label{meson-decomposition}
         \Gamma_\text{M} (q;P)=\sum^N_{i=1}T_i(P;q;\gamma) \,F_i(q^2,q\cdot P,P^2)\;,
         \end{eqnarray}
         where semicolons separate four-vector arguments, and $N=4$ for mesons with total
         spin $J=0$ and $N=8$ otherwise (see e.g., \cite{Krassnigg:2010mh}).
         A detailed account of the $T_i$ for pseudoscalar and vector mesons as needed in our
         calculation can be found in Ref.~\cite{Krassnigg:2009zh}.

         The components are scalar functions of their three scalar arguments: the total momentum
         squared $P^2$, the relative momentum squared $q^2$, and an angular variable $q\cdot P$.
         Note that for an on-shell amplitude $P^2=-M^2$ is fixed, while
         one artificially varies $P^2$ in the solution process of the homogeneous BSE (see e.g., \cite{Blank:2010bp,Blank:2011ph},
         where also all necessary details on the numerical solution method can be found).
         In the corresponding inhomogeneous vertex BSE on the other hand, one would have $P$
         and therefore also $P^2$ as a completely
         independent variable (see, e.g.~\cite{Bhagwat:2007rj,Blank:2010bp,Blank:2010sn}).
         Thus, the on-shell scalar components $F_i(q^2,q\cdot P,P^2)$ effectively depend on the two variables
         $q^2$ and $q\cdot P$. The latter can be parameterized by the variable $z\in [-1,1]$ related to
         the cosine defining the angle between $P$ and $q$. With such a reparameterization in mind,
         the components $F_i$ can be expanded further in Chebyshev polynomials, which
         leaves Chebyshev moments of the $F_i$ as sole functions of $q^2$ (for
         details and an illustration of Chebyshev moments, see \cite{Maris:1997tm,Krassnigg:2003dr}).

         For our case very few Chebyshev moments are sufficient to produce converged
         results. However, in the context of the decay processes we note that,
         as a result of the kinematics in the triangle diagram
         as shown below, one needs to know
         the Bethe-Salpeter amplitudes in a certain region for the relative
         momentum squared $p^2$ in the complex plane. We achieve this by
         a continuation of the relevant Chebyshev polynomials into the complex
         $p^2$ plane via a Taylor-expansion technique up to 4th order which
         yields a converged result.

        \subsection{\label{sec:diquarks}Diquarks}

         The relevance of diquark degrees of freedom in hadron physics has been reviewed in~\cite{Anselmino:1992vg},
         and diquarks are in many respects conceptually
         similar to mesons. In our setup, diquark correlations appear as
         structures in a quark-quark system whose properties can depend on
         the truncation or the effective interaction. In particular, in RL truncation
         diquarks appear as timelike poles in the quark-quark $T-$Matrix, which is an unphysical result
         since diquarks are not color singlets but elements of an antisymmetric
         color antitriplet. This has been identified as a truncation
         artifact, i.e., diquarks disappear from the physical
         spectrum beyond RL truncation~\cite{Bender:1996bb}.
         Nevertheless, the significance of diquark correlations as binding structures \textit{within} baryons
         has become apparent in various baryon form-factor studies, see e.g.~\cite{Cloet:2008re,Nicmorus:2010sd,Eichmann:2010je,Eichmann:2011vu}.
         Moreover, the diquark concept has received support
         from investigations of diquark confinement in Coulomb-gauge
         QCD~\cite{Alkofer:2005ug}.

         In our particular case the diquark BSE in RL truncation
         reads
         \begin{eqnarray}\label{eq:diBSEtr}
         \Gamma_\text{D}(p,P)\,C=-\frac{2}{3}\int\limits_q \mathcal{G}(k^2)\,\frac{T_k^{\mu\nu}}{k^2} \gamma^\mu   \chi_\text{D}(q,P)\,C \,\gamma^\nu \,,
         \end{eqnarray}
         where $C$ is the charge-conjugation matrix, $\Gamma_D$ is the diquark amplitude, $\chi_\text{D}(q;P)\,C=S(q_+) \,\Gamma_\text{D}(q;P)\,C \,S(q_-)$,
         and for all practical purposes the only difference to the meson BSE Eq.~(\ref{eq:bse}) is the color factor.

         In the description of baryons as quark-diquark bound states and also for the description
         of baryonic transitions as given below one needs to know not
         only the diquark amplitudes but also the diquark propagator.
         A defining equation for this propagator can be consistently derived from the
         two-quark Dyson equation and reads schematically~\cite{Eichmann:2009zx}
         \begin{equation}\label{diquark-propagator}
         D^{-1}  = \text{tr}\int\conjg{\Gamma}_D  \,S\,\Gamma_D\,S^T - \text{tr}\int\!\!\!\int \conjg{\Gamma}_D\, K^{-1}\, \Gamma_D \,,
         \end{equation}
         which still contains on-shell amplitudes $\Gamma_D$ resulting from Eq.~(\ref{eq:diBSEtr}). To obtain
         the propagator $D$ for general $P^2$ appropriate ans\"atze for the off-shell amplitudes are chosen~\cite{Eichmann:2009zx}.
         $K$ is the same quark-quark interaction kernel that enters Eqs.~(\ref{eq:bse}--\ref{eq:diBSEtr}),
         and a bar on an amplitude denotes charge conjugation:
         $\conjg{\Gamma}(q,P) = C\,\Gamma(-q,-P)^T C^T$.

        \subsection{\label{sec:baryons}Baryons}

\begin{figure}[t]
                    \begin{center}

                    \includegraphics[scale=0.36]{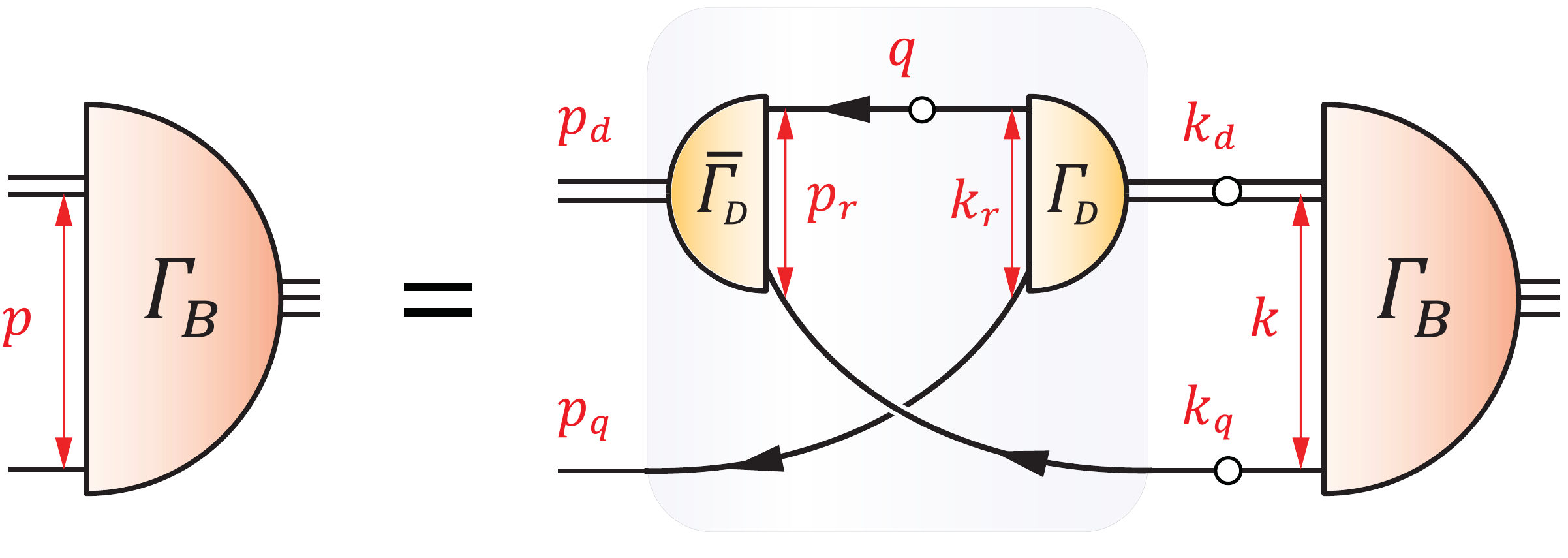}
                    \caption{(Color online) The quark-diquark BSE, Eq.~\eqref{quark-diquark-bse}.\label{fig:barBSE}}

                    \end{center}
        \end{figure}

         In this work baryons are interpreted as bound states of a quark and a
         diquark, which reduces the three-quark problem to an effective two-body problem
         derived via omission of three-quark interactions and a pole ansatz in the
         quark-quark $T$-matrix, see e.g.,
         \cite{Oettel:1998bk,Oettel:2000jj,Eichmann:2007nn,Eichmann:2009zx}. The
         interaction in the resulting equation is an iterative quark exchange,
         where in every iteration step the spectator quark and one quark inside the
         diquark exchange roles. This quark-diquark BSE is illustrated in Fig.~\ref{fig:barBSE} and reads
         \begin{equation}\label{quark-diquark-bse}
          \Gamma_\text{B}^\alpha(p,P)= c^{(\alpha\beta)}\!\!\int\limits_k K_\text{Q-DQ}^{\alpha\beta}\, S(k_q) D^{\beta\beta'}(k_d)\,\Gamma_\text{B}^{\beta'}(k,P)\, ,
         \end{equation}
         where $\Gamma_\text{B}^\alpha$ are the quark-diquark amplitudes of the respective baryon.
         Their general Dirac-Lorentz structure is described e.g. in Refs.~\cite{Oettel:1998bk,Eichmann:2009zx}
         and their decomposition in terms of covariant basis elements and Lorentz-invariant components
         is analogous to the previously discussed case of a meson amplitude, i.e., Eq.~\eqref{meson-decomposition} and below.
         The quark-diquark exchange kernel is given by
         \begin{equation}\label{quark-diquark-kernel}
             K_\text{Q-DQ}^{\alpha\beta} = \Gamma_\text{D}^\beta(k_r,k_d)\,S^T(q)\,\conjg{\Gamma}_\text{D}^\alpha(p_r,p_d)\,,
         \end{equation}
         with $p_{q,d}$ and $k_{q,d}$ being the external and internal quark and diquark momenta and $p_r$, $k_r$
         the relative momenta that enter the diquark amplitudes, cf.~Fig.~\ref{fig:barBSE}.

         The treatment of light baryons such as the nucleon and $\Delta$ in the quark-diquark approach usually
         retains the lightest diquark degrees of freedom, i.e., scalar and axial-vector diquarks.
         The $\Delta$ baryon then involves only axial-vector diquark correlations whereas the nucleon contains both.
         As a consequence, the $\Delta N\pi$ coupling will involve axial-axial contributions as well as axial-scalar diquark transitions.
         The superscripts in Eqs.~(\ref{quark-diquark-bse}--\ref{quark-diquark-kernel})
         account for both kinds of diquarks and an implicit sum over these indices is understood:
         $\Gamma^\alpha$ are scalar ($\alpha=0$) or axial-vector diquark amplitudes ($\alpha=1\dots 4$)
         obtained from their respective diquark BSEs~\eqref{eq:diBSEtr},
         and $D^{00}$ and $D^{\beta\beta'}$ denote the scalar and axial-vector diquark propagators, respectively.
         The color-flavor trace in Eq.~\eqref{quark-diquark-bse} reads $c^{(00)}=-c^{(\alpha\beta)}=-\nicefrac{1}{2}$
         and $c^{(0\alpha)}=c^{(\alpha 0)} = \nicefrac{\sqrt{3}}{2}$ for the nucleon,
         whereas in the case of the $\Delta$ it is given by $c^{(\alpha\beta)}=-1$.
         For details on the solution of the quark-diquark BSE we refer the reader to Refs.~\cite{Oettel:2001kd,Eichmann:2009zx}.

\section{\label{sec:decays}Hadronic decays}

\begin{figure}[t]
                    \begin{center}

                    \includegraphics[scale=0.13]{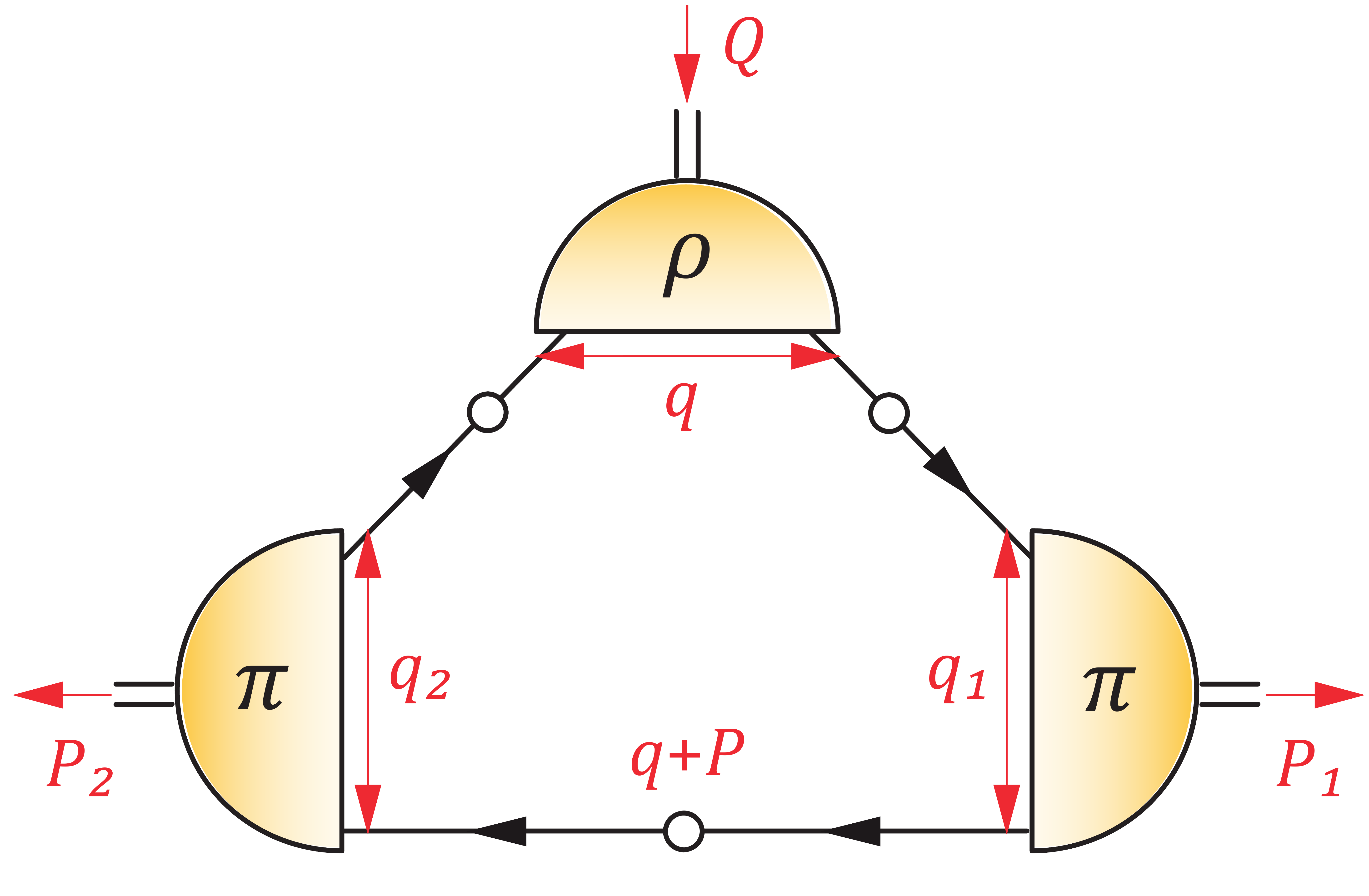}
                    \caption{(Color online) The $\rho\pi\pi$ triangle diagram, Eq.~\eqref{eq:rppvertex}.}\label{fig:triangle}

                    \end{center}
        \end{figure}

         Given a certain truncation of the DSE-BSE system, one can compute
         observables involving various currents through a consistent construction of
         the relevant invariant transition matrix elements. In our case we consider
         transitions between three hadrons and need a quark-level picture of the
         corresponding matrix elements. In the meson case, for RL truncation one
         arrives at a so-called triangle diagram \cite{Jarecke:2002xd}, depicted in
         Fig.~\ref{fig:triangle}, which corresponds to a generalized impulse
         approximation. Analogous diagrams are used at this level for, e.g., meson electromagnetic
         form factors \cite{Maris:1999bh}. For baryons an appropriate construction
         is also possible and given below.

        \subsection{Mesons: $\rho\to\pi\pi$} \label{sec:mesondecays}

          For the meson sector we investigate the $\rho\to\pi\pi$ transition which,
          in terms of Lorentz quantum numbers, corresponds to an $VPP$-vertex.
          As in the usual kinematical setup for a two-body decay one has the total
          $\rho-$meson momentum $Q$ and the relative momentum
          of the pion decay products $P=(P_2-P_1)/2$.
          All mesons are on-shell, i.e. $Q^2=-m_\rho^2$ and $P_1^2=P_2^2=-m_\pi^2$,
          which entails $P\cdot Q=0$.
          Due to the transversality of the $\rho$-meson
         the most general Dirac-Lorentz structure of the transition can be parametrized as
         \begin{equation}
             \Lambda_{\rho\pi\pi}^\mu = 2 P^\mu\,g_{\rho\pi\pi}\,,
         \end{equation}
         where $g_{\rho\pi\pi}$ is its dimensionless coupling constant.
          The corresponding triangle diagram is illustrated in Fig.~\ref{fig:triangle} and reads
         \begin{equation} \label{eq:rppvertex}
             \Lambda_{\rho\pi\pi}^\mu = \text{tr}  \int\limits_q \conjg{\Gamma}_\pi(q_2,P_2)\,S(q+P)\,\conjg{\Gamma}_\pi(q_1,P_1)\,\chi_\rho^\mu(q,Q),
         \end{equation}
         where the $\conjg{\Gamma}_\pi$ are the (charge-conjugated) on-shell pion amplitudes,
         i.e. the canonically normalized solutions of their homogeneous BSEs,
         $S$ is the renormalized dressed quark propagator obtained from its DSE,
         and $\chi_\rho^\mu$ is the $\rho-$meson wave function defined
         in connection with Eq.~\eqref{eq:bse}.
         The traces in color and flavor space yield a factor $3\sqrt{2}$
         in front of the integral, cf. App.~\ref{sec:flavor}.

         More generally, the $\rho\pi\pi$ transition matrix element corresponds to
         the coupling of the pion to an external vector current. Specifically,
         if the $\rho-$meson amplitude that appears in Eq.~\eqref{eq:rppvertex}  were replaced
         by the dressed quark-photon vertex, evaluated at arbitrary momentum squared $Q^2$,
         the respective triangle diagram would constitute the pion's electromagnetic form factor:
         $\Lambda_{\gamma\pi\pi}^\mu = 2 P^\mu F_\pi(Q^2)$.
         As the photon fluctuates into $\rho^0$, the dressed quark-photon vertex, obtained from its inhomogeneous BSE,
         self-consistently develops a $\rho-$meson pole whose on-shell residue
         is proportional to the $\rho-$meson amplitude $\Gamma^\mu_\rho$~\cite{Maris:1999bh}.
         Such a purely transverse component is an important ingredient in
         various hadronic form-factor studies where it contributes typically $\sim 50\%$
         to $\pi$, $N$ and $\Delta$ squared electromagnetic radii~\cite{Maris:1999bh,Bhagwat:2006pu,Eichmann:2008ae,Eichmann:2010je}.
         Consequently, the residue of the pion form factor at the $\rho-$meson pole
         $Q^2=-m_\rho^2$ is proportional to $g_{\rho\pi\pi}$:
         \begin{equation}
             F_\pi(Q^2=-m_\rho^2) \longrightarrow \frac{f_\rho \,m_\rho}{Q^2+m_\rho^2}\,\frac{g_{\rho\pi\pi}}{\sqrt 2}\,.
         \end{equation}

         In the present study we are primarily interested in the value of
         $g_{\rho\pi\pi}$ on the mass shell $Q^2=-m_\rho^2$.
         In a covariant formalism such as ours, one may simply
         choose a frame of reference---in our case the rest frame of the decaying
         particle---by setting $Q=(0,0,0,i m_\rho)$ and $P = (0,0,\kappa,0)$, where
         $\kappa^2 = m_\rho^2/4-m_\pi^2$.
          Then, together with Eq.~(\ref{eq:rppvertex}), $g_{\rho\pi\pi}$ is
          straightforward to evaluate numerically.

         \begin{figure*}[t]
                    \begin{center}

                    \includegraphics[scale=0.11]{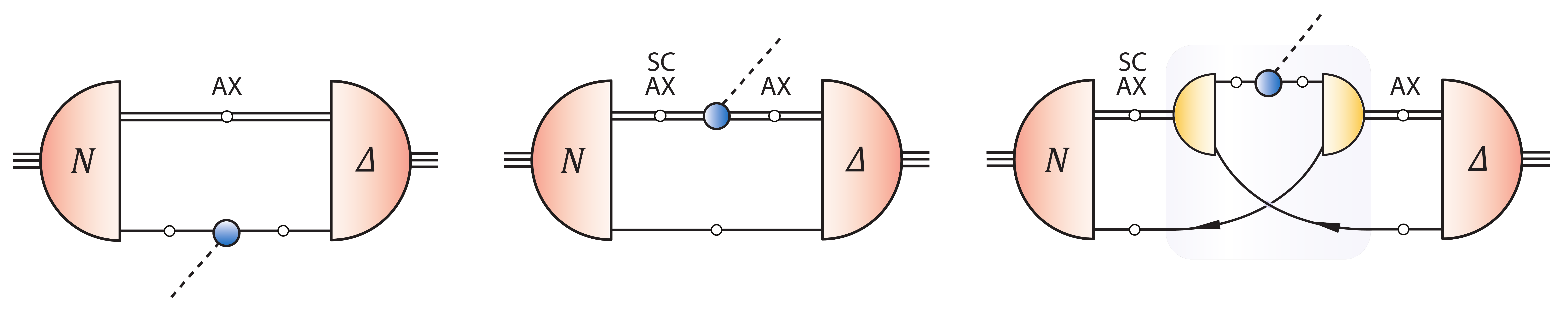}
                    \caption{(Color online) Decomposition of the $\Delta N\pi$ transition matrix element
                             in the quark-diquark model, Eqs.~(\ref{baryon-current}--\ref{current-diagrams}).
                             Seagull terms are neglected. 'SC' and 'AX' denote the types of diquarks (i.e., scalar and axial-vector) that
                             can appear in combination with the nucleon and $\Delta$ bound-state amplitudes.
                             }\label{fig:current}

                    \end{center}
        \end{figure*}

         \subsection{Baryons: $\Delta\to N\pi$} \label{sec:baryondecays}

          In the baryon case, the coupling of an on-shell nucleon and $\Delta-$baryon to a pseudoscalar current
          is described by the pseudoscalar transition form factor $G_{\Delta N\pi}(Q^2)$.
          The $\Delta$-baryon is a spin-3/2 particle and is thus represented by a Rarita-Schwinger spinor.
          Combined with the restriction to positive energies for the nucleon and $\Delta$,
          the most general Dirac-Lorentz structure of the respective interaction vertex is given by
          \begin{equation} \label{eq:LDNpQ}
          \Lambda^\mu_{\text{\tiny{$\Delta N$}}\pi} = G_{\text{\tiny{$\Delta N$}}\pi}(Q^2) \,\frac{Q^\nu}{2 M_N}\,   \Lambda^+_N(P_f)\,\mathbb{P}^{\nu\mu}_\Delta(P_i)\,,
          \end{equation}
          where $P_i$, $P_f$ are the incoming $\Delta$ and outgoing nucleon momenta, $Q=P_f-P_i$ is the pion momentum,
          and $\Lambda^+_B(k) = (\one + \hat{\slashed{k}} )/2$ is the positive-energy
          projector of the baryon with $\hat{k}$ being the respective normalized momentum $\hat{P}_i$ or $\hat{P}_f$.
          The Rarita-Schwinger projector of the $\Delta$ reads
          \begin{equation}
             \mathbb{P}^{\nu\mu}_\Delta(k) = \Lambda^+_\Delta(k) \left(T_k^{\nu\mu}-\frac{1}{3}\,\gamma_T^\nu\gamma_T^\mu\right),
          \end{equation}
          where $T^{\nu\mu}_k$ is a transverse projector with respect to $k$
          and the $\gamma-$matrices $\gamma^\mu_T$ are transverse to $k$ as well.

          At the quark level, the coupling of nucleon and $\Delta$ to a pseudoscalar current is represented by
          the quark-pseudoscalar vertex $\Gamma_5$.
          The latter satisfies an inhomogeneous BSE which has the same structure as Eq.\,\eqref{eq:bse}
          except for an additional inhomogeneous term on the right-hand side~\cite{Maris:1997tm}:
          \begin{equation}\label{inhom-ps-bse}
              \Gamma_5(p,Q) = Z_4\,i\gamma_5 -\frac{4}{3}\int\limits_q \mathcal{G}(k^2)\, \frac{T_k^{\mu\nu}}{k^2} \gamma^\mu   \chi_5(q,Q) \,\gamma^\nu \,.
          \end{equation}
          Its residue at the pion pole $Q^2=-m_\pi^2$ is proportional to the
          homogeneous pion amplitude $\Gamma_\pi$ (cf.~also \cite{Maris:1997tm,Blank:2010sn}):
          \begin{equation}\label{ps-vertex}
               \Gamma_5(p,Q) \longrightarrow  \frac{Z_4}{Z_2}\,\frac{f_\pi m_\pi^2}{2m_0}\,\frac{1}{Q^2+m_\pi^2}\,\Gamma_\pi(p,Q)\,,
          \end{equation}
          where $f_\pi$ is the (calculated) pion decay constant and
          $m_0$ is the bare current-quark mass that enters the quark DSE~\eqref{eq:dse}.
          The solution for $\Gamma_5$ provides an off-shell expression
          for the pion amplitude $\Gamma_\pi$ and thereby allows
          to compute the pseudoscalar transition form factor at spacelike momenta $Q^2>0$.
          Then, $G_{\Delta N\pi}(Q^2)$ corresponds to the form factor obtained from $\Gamma_\pi(p,Q)$ in Eq.~\eqref{ps-vertex},
          where the pion pole as well as its residue are removed, and its on-shell value is
          the $\Delta N\pi$ coupling constant $G_{\Delta N\pi}(-m_\pi^2)=g_{\Delta N\pi}$.

          For the $\Delta\to N\pi$ system the construction analogous to
          Eq.~\eqref{eq:rppvertex} is more complex since there are diquarks
          as well as quarks present as constituents of the states
          involved in the transition. Moreover, the pion will not only interact with the quarks and diquarks directly
          but can couple to the quark-diquark kernel as well,
          i.e. impulse-approximation diagrams alone are no longer sufficient in studying the $\Delta\to N\pi$ transition.

          A systematic procedure to derive the coupling of a baryon to an external current
          is the gauging-of-equations method of Refs.~\cite{Haberzettl:1997jg,Kvinikhidze:1998xn,Kvinikhidze:1999xp}.
          In the context of baryon electromagnetic form factors it has been applied to the quark-diquark model~\cite{Oettel:1999gc}
          as well as the three-quark approach~\cite{Eichmann:2011vu}.
          The starting point is to identify the current with the residue of the 'gauged' quark-diquark (or three-quark)
          $T-$matrix on the baryon's mass shell.
          Upon exploiting the relation between the $T-$matrix and the kernel of the respective bound-state equation,
          the hadronic matrix elements of the current are obtained as a sum of diagrams that describe
          the coupling of the current to all ingredients at the constituent level,
          i.e. in our case to the quark and diquark propagators as well as the quark-diquark kernel.
          The procedure can be applied for mesons as well where, in the case of
          a rainbow-ladder quark-antiquark kernel, the triangle diagram of Fig.~\ref{fig:triangle} is recovered.

         \begin{figure*}[t]
                    \begin{center}

                    \includegraphics[scale=0.33]{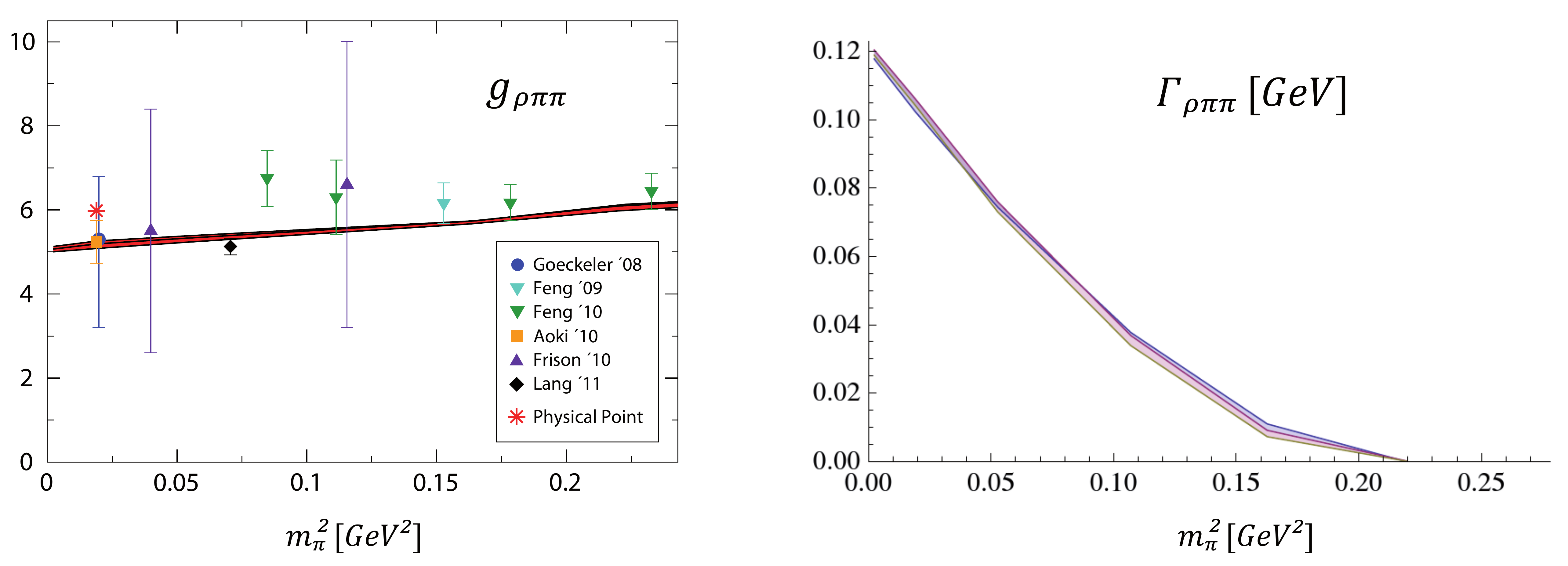}
                    \caption{(Color online)
                              \textit{Left panel:} Evolution of the $\rho\pi\pi$ coupling with the pion mass squared.
                              The experimental point is indicated by the star and
                              the symbols denote lattice data from Refs.~\cite{Gockeler:2008kc,Feng:2009ck,Feng:2010es,Aoki:2010hn,Frison:2010ws,Lang:2011mn}.
                              For better readability the point of the G\"ockeler group is shifted slightly
                              to the right.
                              \textit{Right panel:} The decay width of the $\rho$-meson versus the pion mass squared.
                              The width of the bands illustrates the dependence on $\omega$ (see text).
                             }\label{fig:mesresult}

                    \end{center}
        \end{figure*}

          The generalization of the method from an electromagnetic to a pseudoscalar current, as well as
          different kinds of baryons in the initial and final state, is straightforward.
          The resulting diagrams are displayed in Fig.~\ref{fig:current} and involve
          impulse-approximation couplings to the quarks and diquarks as well
          as a coupling to the exchanged quark that appears in the quark-diquark kernel.
          In principle there would be further diagrams that contain seagulls, i.e., pseudoscalar couplings
          to the diquark amplitudes. For electromagnetic form factors
          such seagull contributions are typically small~\cite{Cloet:2008re,Eichmann:2009zx}
          but necessary to ensure electromagnetic gauge invariance;
          however, in the present case we do not consider them further.
          The $\Delta N\pi$ transition matrix element is then decomposed as
          \begin{equation}\label{baryon-current}
              \Lambda^\mu_{\text{\tiny{$\Delta N$}}\pi} = \Lambda_\text{Q}^\mu + \Lambda_\text{DQ}^\mu + \Lambda_\text{EX}^\mu\,,
          \end{equation}
          where the three contributions are given by
          \begin{equation}\label{current-diagrams}
          \begin{split}
              \Lambda_\text{Q}^\mu  &= \int  \left[ \conjg{\Gamma}_N^\alpha \,\chi_\pi \, \Gamma_\Delta^{\beta\mu}\right] D^{\alpha\beta}\,, \\
              \Lambda_\text{DQ}^\mu &= \int  \left[ \conjg{\Gamma}_N^\alpha \,S\,\Gamma_\Delta^{\beta\mu}\right] D^{\alpha\alpha'} \Gamma_{\text{D}\pi}^{\alpha'\beta'} D^{\beta'\beta}\,, \\
              \Lambda_\text{EX}^\mu &= \int\!\!\!\int  \left[ \conjg{\Gamma}_N^\alpha \,S\,\Gamma_\text{D}^{\beta'}
                                        \chi_\pi^T \,\conjg{\Gamma}_\text{D}^{\alpha'}\,S\, \Gamma_\Delta^{\beta\mu}\right]
                                       D^{\alpha\alpha'}D^{\beta\beta'}\,. \\
          \end{split}
          \end{equation}
          Here we suppressed the explicit momentum dependencies for brevity. The kinematics
          are analogous to the electromagnetic form-factor case and are described, e.g., in App.~C.1 of Ref.~\cite{Nicmorus:2010sd}.
          $\Gamma_N^\alpha$ and $\Gamma_\Delta^{\beta\mu}$ are the quark-diquark amplitudes for the nucleon and $\Delta-$baryon;
          $\chi_\pi = S\,\Gamma_\pi S$ is the pion (off-shell) Bethe-Salpeter wave function
          obtained through the pseudoscalar vertex \eqref{ps-vertex}; $D^{\alpha\beta}$ is the diquark propagator;
          $\Gamma_\text{D}^\alpha$ is the diquark amplitude; and $\Gamma_{\text{D}\pi}^{\alpha\beta}$ is the vertex
          that describes the coupling of the pion to a diquark propagator.
          If an axial-vector diquark is involved, $\alpha,\beta=1\dots 4$ are Lorentz indices.
          Scalar-axialvector transitions, originating from the scalar-diquark component in the nucleon,
          can only occur in $\Lambda_\text{DQ}^\mu$ and $\Lambda_\text{EX}^\mu$;
          in that case: $\alpha=0$, and $D^{00}$ denotes the scalar diquark propagator and $\Gamma_{\text{D}\pi}^{0\beta}$
          the scalar-axialvector transition vertex induced by the pion.

          We note that all ingredients of Eq.~\eqref{current-diagrams} are determined self-consistently:
          Eqs.~\eqref{eq:dse}, (\ref{eq:diBSEtr}--\ref{quark-diquark-bse}) and \eqref{inhom-ps-bse} provide
          the dressed quark propagator, the scalar and axial-vector diquark amplitudes and propagators,
          the baryon amplitudes and the pseudoscalar vertex.
          The diquark-pion vertices are obtained in analogy to Eq.~\eqref{eq:rppvertex}, i.e. through respective triangle diagrams.
          The color and flavor traces in Eq.~\eqref{current-diagrams} are worked out explicitly in App.~\ref{sec:flavor}.

\section{Results and Discussion\label{sec:results}}

        The building blocks of the $\rho\rightarrow\pi\pi$ and $\Delta\rightarrow N\pi$ transition matrix elements
        have been determined in the previous sections and we proceed with computing Eqs.~\eqref{eq:rppvertex} and~\eqref{baryon-current} numerically.
        The Lorentz-invariant coupling constants $g_{\rho\pi\pi}$ and $g_{\text{\tiny{$\Delta N$}}\pi}$ are extracted
        via appropriate momentum contractions and Dirac traces, cf.~Eq.~\eqref{extraction-of-couplings}.
        Since within our chosen truncations the ingredients of the equations for $g_{\rho\pi\pi}$ and $g_{\text{\tiny{$\Delta N$}}\pi}$ are computed selfconsistently,
        the effective quark-gluon coupling defined in Eq.~\eqref{eq:interaction} is the only model parametrization
        with impact upon the results. We explore the sensitivity to this ansatz by varying the $\omega$ parameter
        from its central value $\omega =0.4$ GeV which is indicated by the colored bands in the plots of this section.

        \subsection{Mesons}

         Our result for $g_{\rho\pi\pi}$ at the physical $u/d$-quark mass is shown in Table~\ref{tab:results}
         and compared to the experimental point given by the PDG \cite{Nakamura:2010zzi}.
         In Fig.~\ref{fig:mesresult} we plot $g_{\rho\pi\pi}$
         as a function of the pion mass squared and compare to recent lattice
         results \cite{Gockeler:2008kc,Feng:2009ck,Feng:2010es,Aoki:2010hn,Frison:2010ws,Lang:2011mn}.

         The following observations are important: first, both the magnitude and $m_\pi^2$ dependence
         of our results are in agreement with lattice results, including a slight
         underestimation of the experimental value by about $15\%$. Second,
         the $\omega$ dependence of the result is small, which results from the ground-state
         characteristics of the $\pi$- and $\rho$-meson amplitudes regarding their dependence
         on the relative-momentum squared. This second point also
         solidifies the result of \cite{Jarecke:2002xd} where only the central value of
         our band, $\omega=0.4$ GeV was used.

         Next, the very weak dependence on $m_\pi^2$ compares well to analyses
         using chiral perturbation theory \cite{Hemmert:1997ye,Hanhart:2008mx,Pelaez:2010je}, where one
         concludes from the strong phase-space dependence of the decay width that
         the coupling's $m_\pi^2$-dependence should indeed be close to zero.
         Consequently, when plotting the $\rho\to\pi\pi$ decay width in our approach
         as a function of $m_\pi^2$, Fig.~\ref{fig:mesresult}, the falloff can be almost completely attributed
         to the phase-space factor in Eq.~(\ref{eq:decwidth}) and the width vanishes when
         the decay channel closes.

         \begin{figure*}[t]
                    \begin{center}

                    \includegraphics[scale=0.35]{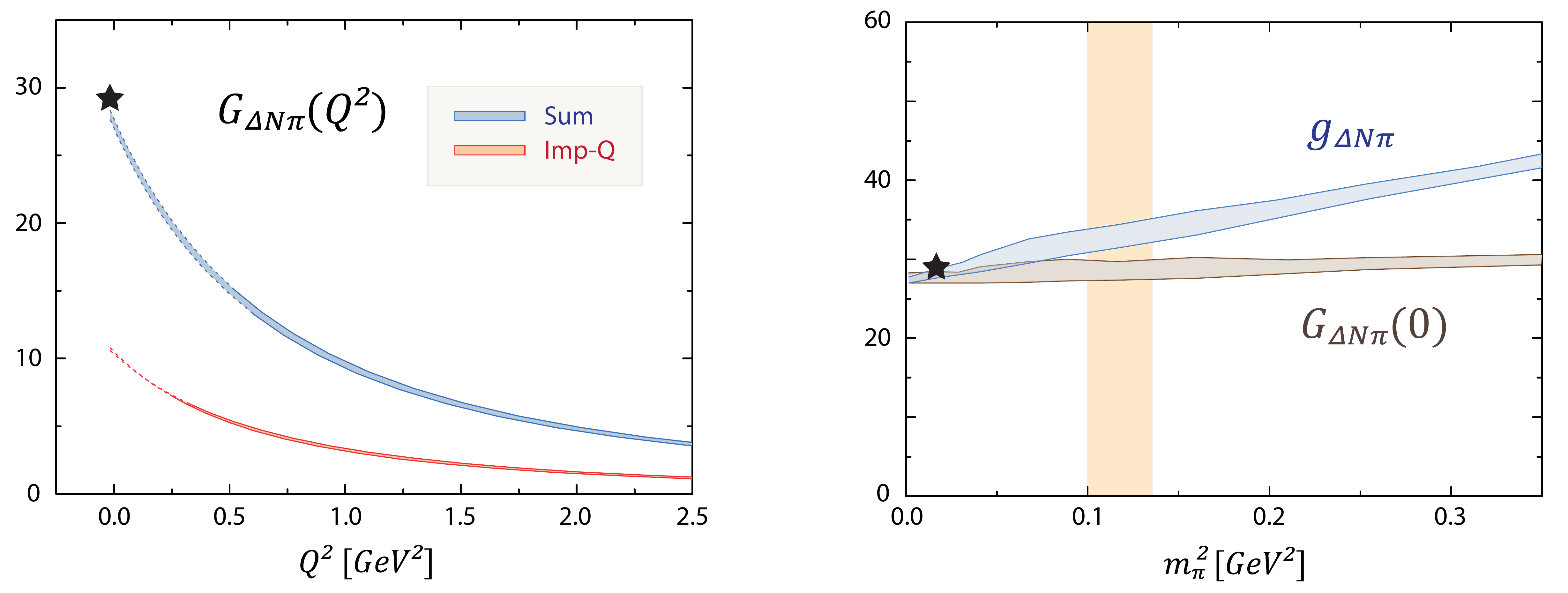}
                    \caption{(Color online)
                              \textit{Left panel:} the transition form factor $G_{\Delta N\pi}$ as a function of the squared pion momentum $Q^2$.
                              The upper band corresponds to the full result and the lower band shows the impulse-approximation value
                              where the pion couples to the quark line only.
                              The width of the bands again shows the dependence on $\omega$.
                              The solid lines are the results of the calculation in the kinematically allowed regions
                              whereas the dashed lines are the respective dipole fits, cf. Eq.~\eqref{dipole}.
                              The pion mass shell $Q^2=-m_\pi^2$ is indicated by the vertical line and the star shows
                              the experimental value $g_{\Delta N\pi} = G_{\Delta N\pi}(-m_\pi^2)$.
                              \textit{Right panel:} Current-mass evolution of $g_{\Delta N\pi}$ (upper band) and $G_{\Delta N\pi}(Q^2=0)$ (lower band).
                              The vertical shaded area depicts the $\omega$-dependent location of the threshold $M_\Delta = M_N + m_\pi$.
                             }\label{fig:barresult}

                    \end{center}
        \end{figure*}

        While comparison to experiment is favorable, the small difference between our result and the
        experimental value for $g_{\rho\pi\pi}$ warrants some discussion.
        The present model calculation uses a simple truncation, but an effective interaction which is
        fitted to the pion mass and underestimates the $\rho$-meson mass by about $5\%$. While
        the resulting kinematical mismatch is responsible for a small part of the difference, the
        main contributions are others. Even though one has to expect some effect from non-resonant corrections
        to RL truncation, the main improvement would be a self-consistent treatment of the $\rho$
        meson as a resonance, i.e., an inclusion of an explicit $\pi\pi$ decay channel in the
        interaction kernel of the vector-meson BSE, in which, of course, also a different (refitted)
        effective interaction would have to be used. However, such an approach is much more involved
        than the present one and clearly beyond the scope of this study, in particular for the baryon case.
        In addition, the reasonably small difference to the experimental value of $g_{\rho\pi\pi}$
        gives reason to expect that for this particular transition the present approach is at least
        a reliable gauge for future studies and results.

        \begin{table}[b]
         \begin{ruledtabular}
         \begin{tabular}{ c @{\;} || @{\;\;}c@{\;\;} | @{\;\;}c@{\;\;} | @{\;\;}c@{\;\;} | @{\;\;}c@{\;\;} || @{\;\;}c@{\;\;}  | @{\;\;}c@{\;\;} }
                                   & $m_\pi$         & $m_\rho$        & $M_N$           & $M_\Delta$       & $g_{\rho\pi\pi}$ & $g_{\text{\tiny{$\Delta N$}}\pi}$ \\ \hline
          This work                & $0.14$          & $0.74$          & $0.94$          & $1.28$           & $5.20$           & $28.1$                            \\ \hline
          Experiment               & $0.14$          & $0.77$          & $0.94$          & $1.23$           & $5.98$           & $29.4$                            \\
          \end{tabular}
         \end{ruledtabular}
          \caption{Comparison of our summarized mass and coupling values at the physical
          point for the central value of the $\omega$ band to corresponding experimental data.
          The pion mass is fitted to experiment, all other numbers are predictions of the
          model with no further parameters adjusted or introduced.
          The masses are given in GeV; the coupling constants are dimensionless. \label{tab:results}}
         \end{table}

       \subsection{Baryons}

       The decay width of the $\Delta-$baryon is governed almost exclusively by the strong interaction,
       namely via the decay into the nucleon and a pion.
       The only other decay channel, the electromagnetic $\Delta\rightarrow N\gamma$ transition,
       has a branching fraction of less than $1\%$.
       Experimentally, $\Gamma_\Delta= 118(2)$~MeV~\cite{Nakamura:2010zzi},
       from which the corresponding coupling strength $g_{\text{\tiny{$\Delta N$}}\pi}=29.4(2)$
       can be inferred via Eq.~\eqref{eq:decwidth}.
        Different conventions that are commonly employed in the literature read
        \begin{equation}
           \frac{g_{\text{\tiny{$\Delta N$}}\pi}}{2M_N} = \frac{g'_{\text{\tiny{$\Delta N$}}\pi}}{m_\pi} = g''_{\text{\tiny{$\Delta N$}}\pi}\,,
           \end{equation}
        with $g'_{\text{\tiny{$\Delta N$}}\pi}\approx 2.16$ and
        $g''_{\text{\tiny{$\Delta N$}}\pi} \approx 15.7$ GeV$^{-1}$.

        For general off-shell momenta the $\Delta\rightarrow N\pi$ coupling is described by the pseudoscalar
        transition form factor $G_{\text{\tiny{$\Delta N$}}\pi}(Q^2)$
        which we obtain from Eqs.~\eqref{baryon-current} and~\eqref{extraction-of-couplings}.
        To compute the $Q^2-$dependence of the form factor we work in the Breit frame
        where the pion momentum is given by $Q=(0,0,|Q|,0)$.
        This has the advantage that the relative momenta in the baryon amplitudes
        are real and no continuation into the complex plane is necessary.
        However, due to the difference in the nucleon and $\Delta$ masses,
        the singularity structure in the quark and diquark propagators
        imposes kinematical constraints on the accessible $Q^2$ region
        from \textit{both} below and above. Hence, to obtain the
        form factor $g_{\text{\tiny{$\Delta N$}}\pi}=G_{\text{\tiny{$\Delta N$}}\pi}(Q^2=-m_\pi^2)$ at the pion mass,
        we fit our results at spacelike $Q^2$ with a dipole form:
        \begin{equation}\label{dipole}
            G_\text{Dipole}(Q^2) = \frac{G_{\text{\tiny{$\Delta N$}}\pi}(0)}{\left(1+Q^2/\Lambda_\pi^2\right)^2},
        \end{equation}
        where $G_{\text{\tiny{$\Delta N$}}\pi}(0)$ and $\Lambda_\pi$ are free fit parameters.

        Our result for the transition form factor $G_{\text{\tiny{$\Delta N$}}\pi}(Q^2)$ at the physical $u/d$ mass
        is shown in the left panel of Fig.~\ref{fig:barresult}.
        Its computed value in the kinematically allowed range is plotted
        as a band with solid margins, where the width of the band
        corresponds to the model parameter $\omega$,
        whereas the fit results in the unaccessible region are shown as dashed lines.
        The resulting value of the strong coupling constant,
        $g_{\text{\tiny{$\Delta N$}}\pi}=28.1$, is remarkably close to the experimental number.
        We also note that the $Q^2-$evolution of $G_{\text{\tiny{$\Delta N$}}\pi}(Q^2)$
        is in good agreement with the lattice data of Ref.~\cite{Alexandrou:2010uk}.

        In general the dipole fits work very well and provide an adequate representation
        of the form factor $G_{\text{\tiny{$\Delta N$}}\pi}$ at spacelike values of the squared pion momentum.
        As described in connection with Eqs.~(\ref{inhom-ps-bse}--\ref{ps-vertex}), the $Q^2-$evolution
        of the form factor is governed by the pseudoscalar vertex $\Gamma_5$
        which includes all pseudoscalar-meson poles, i.e. both the pion's ground state as well as its excitations.
        While the pion ground-state pole and its residue were removed from Eq.~\eqref{ps-vertex}
        to obtain $G_{\text{\tiny{$\Delta N$}}\pi}$, the remaining excited
        states are still encoded in the vertex, hence the form factor
        $G_{\text{\tiny{$\Delta N$}}\pi}$ must diverge at the respective pole locations.
        Indeed we find that the dipole mass $\Lambda_\pi$ in Eq.~\eqref{dipole}
        roughly coincides with the mass of the first excited state in the $0^{-+}$ channel
        which in RL trunctation, at the $u/d$ mass and for the central $\omega$ value,
        is obtained as $m_{\pi^\star} = 1.1$ GeV~\cite{Krassnigg:2004if}.

        The left panel of Fig.~\ref{fig:barresult} also includes the (quark-) impulse-approximation
        contribution to $G_{\text{\tiny{$\Delta N$}}\pi}$,
        i.e. the first diagram in Fig.~\ref{fig:current} corresponding to $\Lambda_\text{Q}^\mu$
        of Eq.~\eqref{baryon-current}. Here the diquark is merely a spectator and,
        since the $\Delta-$baryon only involves axial-vector diquark degrees of freedom,
        only an axial-vector diquark propagator can appear in that diagram.
        Fig.~\ref{fig:barresult} shows that such a direct coupling to the quark
        provides roughly one third of the value of $G_{\text{\tiny{$\Delta N$}}\pi}$.
        The axial-axial contributions stemming from the second and third diagrams are
        comparatively small and contribute $\sim 10\%$ to the full result.
        The remainder owes in equal parts to the axial-scalar transitions that are
        generated from the transition vertex $\Gamma_{\text{D}\pi}^{0\beta}$
        in the second diagram and the axial-scalar contribution
        $\Gamma_\text{D}^\beta \, \chi_\pi^T\,\conjg{\Gamma}_\text{D}^0$ in the exchange diagram,
        cf.~Eq.~\eqref{current-diagrams}.

        Once again, the current-quark mass dependence of the transition form factor
        can be studied by varying the current mass in the quark DSE.
        That change will be reflected in all ingredients that enter the transition matrix element.
        Similarly to the meson case, the overwhelming contribution to the mass dependence
        of the decay width $\Gamma_{\text{\tiny{$\Delta N$}}\pi}$ comes from the
        phase-space factor in Eq.~\eqref{eq:decwidth}.
        This is especially conspicuous in the form factor
        $G_{\text{\tiny{$\Delta N$}}\pi}(Q^2=0)$ at vanishing pion momentum,
        shown in the right panel of Fig.~\ref{fig:barresult},
        which is practically independent of the current-quark mass.
        Similar features have been reported for $N$ and $\Delta$
        electromagnetic form factors~\cite{Nicmorus:2010sd,Eichmann:2011vu}.
        The observation stays true for the $Q^2-$evolution, i.e.
        $G_{\text{\tiny{$\Delta N$}}\pi}(Q^2)$ as well as its individual contributions
        retain their shape throughout the current-mass range
        if they are plotted over a dimensionless variable such as $Q^2/M_N^2$.
        Considering Eq.~\eqref{dipole}, this means that the dipole fit works also well
        for higher quark masses since the mass of the excited pion also varies with the
        current-quark mass in a similar fashion as $M_N$ and $M_\Delta$.

        On the other hand, the value of
        $g_{\text{\tiny{$\Delta N$}}\pi} = G_{\text{\tiny{$\Delta N$}}\pi}(-m_\pi^2)$ rises with the
        quark mass because of the current-mass dependent pion pole location.
        This property is also visible for $g_{\rho\pi\pi}$ in Fig.~\ref{fig:mesresult},
        albeit less pronounced, as the $\rho-$meson mass
        is non-zero in the chiral limit and therefore
        varies over a much smaller range when evolving the current-quark mass.
        The shaded area in Fig.~\ref{fig:barresult} indicates the threshold position
        $M_\Delta = M_N+m_\pi$ where the decay channel closes,
        and the width of the band is again induced by the $\omega-$dependence
        which enters mainly through the mass of the $\Delta$, cf.~Ref.~\cite{Nicmorus:2010mc}.

        We finally note that in determining the $\Delta\to N\pi$ transition form factor
        we have neglected the pseudoscalar seagull terms which would appear in addition
        to the diagrams displayed in Fig.~\ref{fig:barresult}. Judging from the smallness
        of the electromagnetic seagulls in the case of electromagnetic form factors, this
        approximation might be well justified. The question of its validity can be settled
        by investigating the $\Delta\to N\pi$ transition in the three-quark framework of
        Ref.~\cite{Eichmann:2011vu} where all such missing contributions, while no longer
        appearing explicitly, would be automatically included.

\section{\label{sec:conclusions}Conclusions}

        We presented a calculation of the hadronic $\rho\to\pi\pi$ and $\Delta\rightarrow N\pi$ decays,
        as well as the pseudoscalar transition form factor $G_{\text{\tiny{$\Delta N$}}\pi}(Q^2)$,
        in the framework of Dyson-Schwinger and covariant bound-state equations.
        A consistent construction for the decay diagrams was implemented.
        The $\rho\to\pi\pi$ transition was computed from the quark-antiquark
        Bethe-Salpeter equation in rainbow-ladder truncation whereas
        the $\Delta\to N\pi$ transition was studied within the covariant quark-diquark model.
        All ingredients are determined self-consistently which leaves a
        phenomenological ansatz for the quark-gluon coupling as the only model input.

        The results in both cases compare well with experimental and lattice data.
        The $\rho\pi\pi$ coupling is underestimated by $\sim 15\%$ and slowly rises with the current-quark mass,
        in agreement with lattice results.
        A similar observation holds for the $\Delta\to N\pi$ coupling which agrees also well with the experimental result.
        We find that $G_{\text{\tiny{$\Delta N$}}\pi}(0)$ is practically independent of the current-quark mass.
        Consequently, the decay widths for $\rho$ and $\Delta$ are mainly governed by the available phase space.

        The present calculation provides a first step towards a thorough investigation of hadron resonances and their decays within QCD.
        An important future direction in that respect would involve the implementation of explicit $\rho\pi\pi$ and $\Delta N\pi$ decay channels
        in the $\rho-$meson and $\Delta-$baryon bound-state equations.
        Such an extension would represent a considerable step beyond the rainbow-ladder truncation employed herein
        and contribute to a more realistic description of hadron resonances from their underlying dynamics in QCD.

       \section*{Acknowledgments}

       We acknowledge valuable discussions with R.~Alkofer, C.~S.~Fischer, D.~Nicmorus and R.~Williams.
       This work was supported by the Austrian Science Fund \emph{FWF} under Project No.\ P20496-N16,
       Doctoral Program no.\ W1203-N08, and Erwin-Schrodinger-Stipendium No.\ J3039, as well as the
       Helmholtz International Center for FAIR within the
       framework of the LOEWE program launched by the
       State of Hesse, GSI, BMBF and DESY.

\begin{appendix}

        \section{Matrix elements and decay widths} \label{sec:currents}

         For the decay of a particle with momentum $p$ and mass $M$ into two decay products with momenta $p_i$ and masses $m_i$,
         with $p=\sum_i p_i$,
         the decay width is given by
         \begin{equation}\label{M's}
             \Gamma = \frac{I}{2M}\,|\mathcal{M}|^2\,,
         \end{equation}
         where the squared transition matrix element $|\mathcal{M}|^2$ is averaged over the spins/polarizations, i.e.
         \begin{equation}
         \begin{split}
             |\mathcal{M}_{\rho\pi\pi}|^2& = \frac{1}{3} \sum_\lambda |\mathcal{M}_\lambda|^2\,, \\
             |\mathcal{M}_{\Delta N\pi}|^2 &= \frac{1}{4} \sum_{ss'} |\mathcal{M}_{ss'}|^2
         \end{split}
         \end{equation}
         and a sum over all final states is implicit.
         The phase-space factor $I=\kappa/(4\pi M)$ involves the quantity
         \begin{equation}\label{eq:kappa}
             \kappa = \frac{\sqrt{\left(M^2-(m_1+m_2)^2\right)\left(M^2-(m_1-m_2)^2\right)}}{2M}
         \end{equation}
         which in the rest frame of the decaying particle is given by $\kappa=|\vect{p_1}| = |\vect{p_2}|$;
         thus the decay width becomes
         \begin{equation}
             \Gamma = \frac{\kappa}{8\pi M^2}\,|\mathcal{M}|^2\,.
         \end{equation}

         In the case $\rho\rightarrow\pi\pi$ one obtains specifically:
         \begin{equation}
             \mathcal{M}_\lambda = \Lambda^\mu_{\rho\pi\pi}\,\epsilon^\mu_\lambda \quad \Longrightarrow \quad
             |\mathcal{M}_{\rho\pi\pi}|^2 = \frac{4}{3}\,\kappa_{\rho\pi\pi}^2 \,g_{\rho\pi\pi}\,,
         \end{equation}
         where from Eq.~\eqref{eq:kappa} and the kinematics in Fig.~\ref{fig:triangle} one has
         \begin{equation}
             \kappa_{\rho\pi\pi} = \sqrt{m_\rho^2/4-m_\pi^2} = \sqrt{P^2}\,,
         \end{equation}
         and the polarization vectors $\varepsilon_{\lambda}^\mu$ of the $\rho$-meson are normalized to
         $\varepsilon_{\lambda}^\mu\,\varepsilon_{\lambda'}^\mu = \delta_{\lambda\lambda'}$.
         In the case $\Delta\rightarrow N\pi$ Eq.~\eqref{eq:LDNpQ} entails
         \begin{equation}
         \begin{split}
             \mathcal{M}_{ss'} &= \conjg{\sp}_s \, \Lambda^\mu_{\text{\tiny{$\Delta N$}}\pi}  \,\sp^\mu_{s'} =
                                  \frac{g_{\text{\tiny{$\Delta N$}}\pi}}{2M_N}\,\conjg{\sp}_s  \,Q^\mu \,  \sp^\mu_{s'}\,,
         \end{split}
         \end{equation}
         where $Q$ is the pion momentum, and the nucleon and $\Delta$ spinors are normalized to
         \begin{equation}
             \sum_s \sp_s \conjg{\sp}_s = 2 M_N \,\Lambda_N^+\,, \quad
             \sum_s \sp_s^\mu \,\conjg{\sp}_s^\nu = 2 M_\Delta \,\mathbb{P}^{\mu\nu}_\Delta\,,
         \end{equation}
         respectively.
         This yields
         \begin{equation}
            |\mathcal{M}_{\text{\tiny{$\Delta N$}}\pi}|^2 = \frac{1}{6} \frac{M_\Delta}{M_N}\,\sigma \,\kappa_{\text{\tiny{$\Delta N$}}\pi}^2 \,g_{\text{\tiny{$\Delta N$}}\pi}^2
         \end{equation}
         with the spin sum
         \begin{equation}
             \sigma = \text{tr}\left\{ \Lambda_N^+(P_f)\,\Lambda_\Delta^+(P_i) \right\} = 1 + \sqrt{1+\frac{\kappa_{\text{\tiny{$\Delta N$}}\pi}^2}{M_\Delta^2}}\,.
         \end{equation}
         The strong decay widths in both cases are then given by 
         \begin{equation}\label{eq:decwidth}
             \Gamma_{\rho\pi\pi} = \frac{\kappa_{\rho\pi\pi}^3 \,g_{\rho\pi\pi}^2}{6\pi m_\rho^2}\,, \quad
             \Gamma_{\text{\tiny{$\Delta N$}}\pi} = \frac{\sigma \,\kappa_{\text{\tiny{$\Delta N$}}\pi}^3 \,g_{\text{\tiny{$\Delta N$}}\pi}^2}{48\pi M_\Delta M_N}\,.
         \end{equation}

         To extract the coupling constants from the matrix elements in Eqs.~\eqref{eq:rppvertex} and~\eqref{baryon-current},
         one has to perform appropriate momentum contractions and Dirac traces which yields:
         \begin{equation}\label{extraction-of-couplings}
         \begin{split}
             g_{\rho\pi\pi} = \frac{P^\mu \Lambda_{\rho\pi\pi}^\mu}{2\kappa_{\rho\pi\pi}^2}\,, \quad
             g_{\text{\tiny{$\Delta N$}}\pi} = \frac{3M_N}{\sigma \kappa_{\text{\tiny{$\Delta N$}}\pi}^2}\,\text{tr} \left\{Q^\mu \Lambda^\mu_{\text{\tiny{$\Delta N$}}\pi} \right\}\,.
         \end{split}
         \end{equation}

\section{Color and flavor factors} \label{sec:flavor}

        While the color and flavor traces in the quark DSE~\eqref{eq:dse}
        and meson and baryon bound-state equations, Eqs.~(\ref{eq:bse},\,\ref{eq:diBSEtr},\,\ref{quark-diquark-bse}), have
        already been worked out in the main text, we still have to perform these traces
        for the $\rho\rightarrow\pi\pi$ and $\Delta\rightarrow N\pi$
        transition matrix elements~\eqref{eq:rppvertex} and~\eqref{baryon-current}.

        We work in the $SU(2)_f$-isosymmetric limit and thus we have two
        degenerate quark flavors  $u=\bigl( \begin{smallmatrix}1\\0 \end{smallmatrix}\bigr)$ and
        $d=\bigl( \begin{smallmatrix}0\\1 \end{smallmatrix}\bigr)$. They transform according
        to the fundamental representation of $SU(2)$, whereas the anti-quarks
        $\bar u=\bigl( \begin{smallmatrix}1\\0 \end{smallmatrix}\bigr)$ and
        $\bar d=\bigl( \begin{smallmatrix}0\\1 \end{smallmatrix}\bigr)$ transform
        according to the complex conjugated fundamental representation. These can be used
        to construct representation matrices of mesons as quark-antiquark, diquarks
        as quark-quark and baryons as quark-diquark bound states.

        The $\pi$ and the $\rho$-mesons discussed in this article are isovector
        states with three isospin projections, which can be labeled by the
        corresponding electric meson charge. A possible
        representation is given by the matrices
        \begin{equation}\label{eq:mesflavmatr}
         \begin{split}
           &\mathsf{r_+} = \ket{u\,\bar d} = \frac{\sigma_1 + i \sigma_2}{2}\,, \\
           &\mathsf{r_0} \,= \frac{ \ket{u\,\bar u}-\ket{d\,\bar d} }{\sqrt{2}} =   \frac{\sigma_3}{\sqrt 2 } \,, \\
           &\mathsf{r_-} =   \ket{d\,\bar u} =  \frac{\sigma_1 - i \sigma_2}{2}\,,
         \end{split}
        \end{equation}
        where the $\sigma_i$ are the Pauli matrices, and the flavor
        matrices are normalized to $\text{tr}\{\mathsf{r}_i^\dag\,\mathsf{r}_j^\dag\}=\delta_{ij}$.
        The color factors for the mesons are given by $\delta_{AB}$, where $A,B=1,2,3$ denote the quark indices.

        The $\rho^0-$meson in the triangle diagram of Eq.~\eqref{eq:rppvertex} can couple to the upper
        and the lower quark line. The color trace in both diagrams equals $3$ whereas the
        Dirac traces yield an opposite sign: $(\Lambda^\uparrow)_{\rho\pi\pi}^\mu = -(\Lambda^\downarrow)_{\rho\pi\pi}^\mu$,
        where $\Lambda^\uparrow$ represents the expression in Eq.~\eqref{eq:rppvertex}.
        The full color-flavor-Dirac trace of the $\rho^0\to\pi^+\pi^-$ transition then yields
        \begin{equation*}
            -\text{tr}\{\mathsf{r}_+^\dag \,\mathsf{r}_-^\dag \,\mathsf{r}_0\} \,3 \,\Lambda^\uparrow -
            \text{tr}\{\mathsf{r}_+^\dag \,\mathsf{r}_0 \,\mathsf{r}_-^\dag \} \,3 \,\Lambda^\downarrow = 3\sqrt{2}  \, \Lambda^\uparrow\,,
        \end{equation*}
        i.e. the color-flavor trace of Eq.~\eqref{eq:rppvertex} is $3\sqrt{2}$.

        Similar to the mesonic case, for two degenerate flavors there exist four different diquarks,
        one of which is an isoscalar whereas the other three form an iso-triplet. The difference to
        mesons in flavor space is due to the different transformation properties of quarks and
        antiquarks. The representation for the diquark flavor matrices reads
        \begin{equation}\label{eq:diqflavmatr}
        \begin{split}
          \mathsf{s_0} &= \frac{\ket{u\,d} - \ket{d\,u}}{\sqrt 2} = \frac{i\sigma_2}{\sqrt{2}}\,, \\
          \mathsf{s_1} &= \ket{u\,u} = \frac{\one + \sigma_3 }{2}\,,\\
          \mathsf{s_2} &= \frac{\ket{u\,d} + \ket{d\,u}}{\sqrt 2} = \frac{\sigma_1}{\sqrt{2}}\,,\\
          \mathsf{s_3} &= \ket{d\,d} = \frac{\one - \sigma_3}{2} \,.
        \end{split}
        \end{equation}

        The flavor factors for baryons in the quark-diquark picture are given by the
        Clebsch-Gordan coefficients according to the respective diquark content of the
        baryon. For proton and neutron they read
        \begin{equation} \label{eq:CGN-nuc}
        \begin{split}
         \mathsf{p} &= \left(\,u \;\Big| \,\sqrt{\tfrac{ 2}{3}} \,d\,,\, -\sqrt{\tfrac{1}{3}}\, u\,,\,0\, \right),\\
         \mathsf{n} &= \left(\,d \;\Big| \,0\,,\,\sqrt{\tfrac{ 1}{3}} \,d\,,\, -\sqrt{\tfrac{2}{3}}\, u\, \right) ,
         \end{split}
        \end{equation}
        where the first terms represent the isoscalar diquark contributions and the remaining
        three the contributions from the isovector channel in the same order as Eq.~\eqref{eq:diqflavmatr}.
        The $\Delta$-baryons do not contain any contribution from the scalar diquark, and
        the corresponding Clebsch-Gordan construction yields
        \begin{equation} \label{eq:CGN-delta}
        \begin{split}
         &\!\Delta^{++} = \Big(\,u\,, \,0\,,\,0\, \Big) , \quad \Delta^{+}= \left(\,\sqrt{\tfrac{ 1}{3}} \,d\,,\,\sqrt{\tfrac{2}{3}}\, u\,,\,0\, \right), \\
         &\!\Delta^{0} = \left(\,0\,,\,\sqrt{\tfrac{2}{3}} \,d\,,\,\sqrt{\tfrac{1}{3}}\, u\, \right) , \quad \Delta^{-} = \Bigl(\,0,\,0,\,d\, \Bigr).
        \end{split}
        \end{equation}
        Finally, the color factors are $(\varepsilon_{ABC}/\sqrt{6})$ for the diquark amplitudes
        and $(\delta_{AC}/\sqrt{3})$ for the $N$ and $\Delta$ quark-diquark amplitudes,
        where $A,B$ are quark indices and $C$ is the diquark index.

        In the case of the $\Delta^{++}\to p\pi^+$ transition the three contributions in Eq.~\eqref{current-diagrams}
        yield the following flavor traces:
        \begin{equation}
        \begin{split}
            \sum_j \mathsf{p}_j^\dag \,\mathsf{r}_+^\dag \,\Delta^{++}_j &\,, \quad
            \sum_{ij} \mathsf{p}_i^\dag \, \Delta^{++}_j\,2\,\text{tr}\{ \mathsf{s}_i^\dag\,\mathsf{s}_j\,\mathsf{r}_+^\dag \}, \\
            &\sum_{ij} \mathsf{p}_i^\dag \, \mathsf{s}_j\,\mathsf{r}_+^\dag\,\mathsf{s}_i^\dag\,\Delta^{++}_j\,.
        \end{split}
        \end{equation}
        Only the axial-vector diquark contributes to the $\Delta$ (hence $j=1,2,3$)
        whereas the nucleon has both scalar ($i=0$) and axial-vector diquark components ($i=1,2,3$).
        Combined with the color traces $+1$ for the impulse-approximation diagrams and $-1$ for the
        exchange diagrams the final result for the $\Delta^{++}\to p\pi^+$ transition matrix element in Eq.~\eqref{baryon-current} reads
        \begin{equation*} \label{flavor-final}
        \Lambda = \textstyle{\sqrt{\frac{2}{3}}}\left[ \Lambda_\text{Q}^\text{AA} - \Lambda_\text{DQ}^\text{AA} +
                  \sqrt{3}\,\Lambda_\text{DQ}^\text{SA} + \textstyle{\frac{1}{2}}\,\Lambda_\text{EX}^\text{AA} - \textstyle{\frac{\sqrt{3}}{2}}\,\Lambda_\text{EX}^\text{SA} \right],
        \end{equation*}
        where the superscripts S and A refer to the scalar or axial-vector
        diquark content in the outgoing (left) and incoming (right) baryon amplitudes.

        The remaining processes are obtained accordingly by replacing $\mathsf{p}$, $\Delta^{++}$ and $\mathsf{r_+}$ with the appropriate
        flavor structures of Eqs.~\eqref{eq:mesflavmatr} and (\ref{eq:CGN-nuc}--\ref{eq:CGN-delta}).
        The bracket in the previous equation is identical in all cases whereas
        the prefactors become:
        \begin{itemize}
        \item $\pm\sqrt{\frac{2}{3}}$ for the transitions $\Delta^{++}\to p\pi^+$, $\Delta^-\to n\pi^-$;
        \item $-\frac{2}{3}$ for $\Delta^+\to p\pi^0$, $\Delta^0\to n\pi^0$;
        \item $\pm\frac{\sqrt{2}}{3}$ for $\Delta^+\to n\pi^+ $ and $\Delta^0\to p\pi^-$.

        \end{itemize}
        For every initial state one has to sum over all final states via Eq.~\eqref{M's}, hence one gets
        for both $\Delta^+$ and $\Delta^0$
        \be
        \sqrt{\abs{\frac 2 3}^2+\abs{\frac{\sqrt 2}{3}}^2} = \sqrt{\frac 2 3}\,,
        \ee
        and thus all flavor factors in the $\Delta N\pi$ system are the same.

\end{appendix}


%

\end{document}